\documentclass[%
 aip,
 amsmath,amssymb,
reprint,%
]{revtex4-1}

\usepackage{graphicx}
\usepackage{dcolumn}
\usepackage{bm}

\usepackage[utf8]{inputenc}
\usepackage[T1]{fontenc}
\usepackage{mathptmx}

\graphicspath{{Figures/}}
\usepackage{xcolor}
\usepackage[normalem]{ulem}

\begin{document}

\preprint{AIP/123-QED}

\title{Sub-millimetric ultra-low-field MRI \\detected \textit{in situ} by a dressed atomic magnetometer}

\author{Giuseppe Bevilacqua}
 \author{Valerio Biancalana}%
 \email{valerio.biancalana@unisi.it}
\affiliation{
 DIISM - University of Siena  \\ Via Roma 56 Siena Italy
}

\author{Yordanka Dancheva}\altaffiliation[currently at: ]{Aerospazio Tecnologie S.r.l., Via Provinciale Nord, 42a Rapolano, Siena (Italy)}
\affiliation{
 DSFTA - University of Siena  \\ Via Roma 56 Siena Italy
}

\author{Antonio Vigilante}
\affiliation{
 DSFTA - University of Siena  \\ Via Roma 56 Siena Italy
}

\date{\today}
                              
\begin{abstract}
Magnetic Resonance Imaging (MRI) is universally acknowledged as an excellent tool to extract detailed spatial information with minimally invasive measurements. Efforts toward ultra-low-field (ULF) MRI are made to simplify the scanners and to reduce artefacts and incompatibilities. Optical Atomic Magnetometers (OAMs) are among the sensitive magnetic detectors eligible for ULF operation, however they are not compatible with the strong field gradients used in MRI. We show that a magnetic-dressing technique restores the OAMs operability despite the gradient, and we demonstrate sub-millimetric resolution MRI with a compact experimental setup based on an \textit{in situ} detection.  The proof-of-concept experiment produces unidimensional imaging of remotely magnetized samples with a dual sensor, but the approach is suited to be adapted for 3-D imaging of samples magnetized \textit{in loco}. An extension to multi-sensor architectures is also possible.

\end{abstract}

\maketitle

Isidor Rabi  was awarded the Nobel Prize in Physics in 1944 for his seminal research, which, in 1938, demonstrated the phenomenon of nuclear magnetic resonance (NMR) in a molecular beam \cite{rabi_pr_38}. Felix Bloch and Edward Purcell were awarded the Noble Prize for their independent contributions to NMR (dated 1946) \cite{bloch_pr_46, purcell_pr_46} in 1952, the same year in which Robert Gabillard,  in his PhD thesis, studied the NMR in the presence of magnetic field gradients (a crucial step in view of encoding spatial information in the precessing nuclei). 
It took two additional decades to realize that the potential of NMR to record spatial distribution of precessing nuclei could be exploited as a medical diagnostic tool, as proposed by Raymond Damadian \cite{damadian_science_71}. The applicability of that idea was then demonstrated by Peter Mansfield and Paul Lauterbur \cite{lauterbur_nat_73}, which  thirty years after were awarded the Nobel Prize.

The attractiveness of MRI in medicine relies on its accurateness and on its substantially non-invasive nature. The latter feature is shared with ultrasonography \cite{merritt_rad_89}, whose  development occurred almost simultaneously with MRI. Both methodologies are spreadly used and constitute favourite choices with respect to more invasive imaging techniques based on ionizing radiation.

The great impulse impressed to the development of the MRI technology,  led to fast and impressive progresses in the methodologies used to generate,  detect, and  analyze  MRI signals. These advances were facilitated by the parallel progresses achieved in some related technologies, such as electronics, computer science  and cryogenics.

Most of MRI and, more generally, NMR advances followed the straightforward direction of enhancing strength and homogeneity of the magnetic field (or accurate control of its gradients) as well as increasing the signal-to-noise ratio (SNR) of the detection stage. Cryogenics and superconductor technologies constituted an obvious opportunity for this evolution.

At the same time, cryogenics
--allowing the development of innovative detectors (superconducting quantum interference devices, SQUIDs) with unrivaled sensitivity-- made available alternative (non inductive) sensors, and opened the perspective of performing NMR and MRI at much lower precession frequencies, that is at low and ultra-low field  strengths \cite{kraus_book_14}.

The ULF-NMR  dates about three decades, similarly to ULF-MRI \cite{sepponen_jcat_85}, for which intense progresses started, however, less than two decades ago \cite{mcdermott_pnas_04, moessle_ieee_05}.

MRI in the ULF regime  comes with several valuable advantages \cite{kraus_book_14}.  
The ultimate spatial resolution of MRI is determined by the NMR spectral resolution, that depends on the \textit{ absolute} field inhomogeneity. At ULF, a modest \textit{ relative} field homogeneity  turns out to be excellent on an absolute scale: very narrow NMR lines with a high SNR can be recorded in ULF regime, using relatively simple and unexpensive field generators \cite{mcdermott_sc_02, matlachov_jmr_04, burghoff_apl_05}. The encoding gradients necessary for MRI can be generated by simple and low-power coil systems, as well \cite{mcdermott_pnas_04, zotev_ieee_07}. Further important advantages 
brougth to MRI by ULF regime include
the  minimization of susceptibility \cite{seung_mrm_04} and conductance \cite{matlachov_jmr_04, moessle_jmr_06} artefacts.
Other delicate instrumentation (not compatible with strong and/or fast-varying magnetic fields) can be used in conjunction with ULF-MRI in more complex setups. In addition, the non-conventional magnetic detectors used in ULF-MRI can be used to record low-frequency magnetic signals originating not only from the precessing nuclei, but also from other (e.g. biomagnetic) sources.  Hybrid instrumentation enabling multimodal MRI and magneto-encephalo-graphic measurements in medical applications has been demonstrated \cite{panu_mrm_12,matlachov_ieee_05} 
(see also Chap.5 in Ref.\cite{kraus_book_14}).

While in conventional NMR and MRI, the  premagnetization and precession are typically induced by one field, the two functions are often distinguished in ULF apparatuses. Here, the precession field can be extremely weak, and the homogeneity of the (strong) premagnetization field is not a critical parameter. It is worth mentioning that  schemes of no-magnet NMR, with zero precession field and alternative premagnetization methods have been proposed \cite{theis_natphys_11}.

Direct coupling of  SQUID sensors with strong
premagnetization fields in not feasible.  
Successful attempts to overcome this problem are based on using flux dams \cite{koch_apl_95} or finely
designed SQUID coils, with gradiometric sensitivity and very high
rejection of the common mode term \cite{matlachov_ieee_05}, while another possible approach is based on using OAMs as
alternative highly sensitive non-inductive detectors \cite{tayler_rsi_17}.
Beside robustness, OAMs bring the advantage of not requiring cryogenics, so to be a favourite choice whenever ULF systems are designed in view of building up simpler and low-cost apparatuses. Despite their simplicity, OAMs --in some implementations-- may compete with SQUIDs in terms of sensitivity. In facts, the literature reports successful ULF-MRI experiments using both SQUIDs and OAMs as highly sensitive, non-inductive sensors \cite{zotev_sst_07, savukov_jmr_13}. 

OAMs operate on the basis of paramagnetic atoms in which an atomic magnetic resonance (AMR) is induced using   resonant light as a polarization tool (modern laser spectroscopy methodologies provide very effective instrumentation to this end)
\cite{budker_nat_07}.
The sensitivity of OAMs relies on the narrow spectral width of the AMR. The important field gradients necessary for MRI applications would broaden severely the AMR. As matter of fact, the ULF-MRI 
experiments reported
so far with OAM detection are based on \textit{ex-situ} measurements: the magnetic signal produced by the precessing nuclei is coupled to the sensor via flux transformers \cite{savukov_apl_13}, eventually resulting in remote-detection techniques \cite{xu_pnas_06}.

In this letter we demonstrate that an approach based on an inhomogeneous magnetic dressing of the precessing atoms \cite{biancalana_IDEA_prappl_19} can be used to record \textit{in situ} MRI signals by means of OAM, achieving sub-millimetric resolution. The described proof-of-concept experiment makes use of a dual sensor, but paves the way to multi-sensor detection, with the potential of improving the spatial resolution, enhancing the allowed sample size, and speeding up the acquisition. 

The described setup performs MRI of samples premagnetized in strong field and subsequently transferred to the detection region, however the robustness of OAMs to strong magnetic fields would enable MRI of samples premagnetized in the same position where the NMR signal is detected, so to obtain a fully static operation.

\begin{figure}[ht]
\includegraphics[angle=-90,width= \columnwidth]{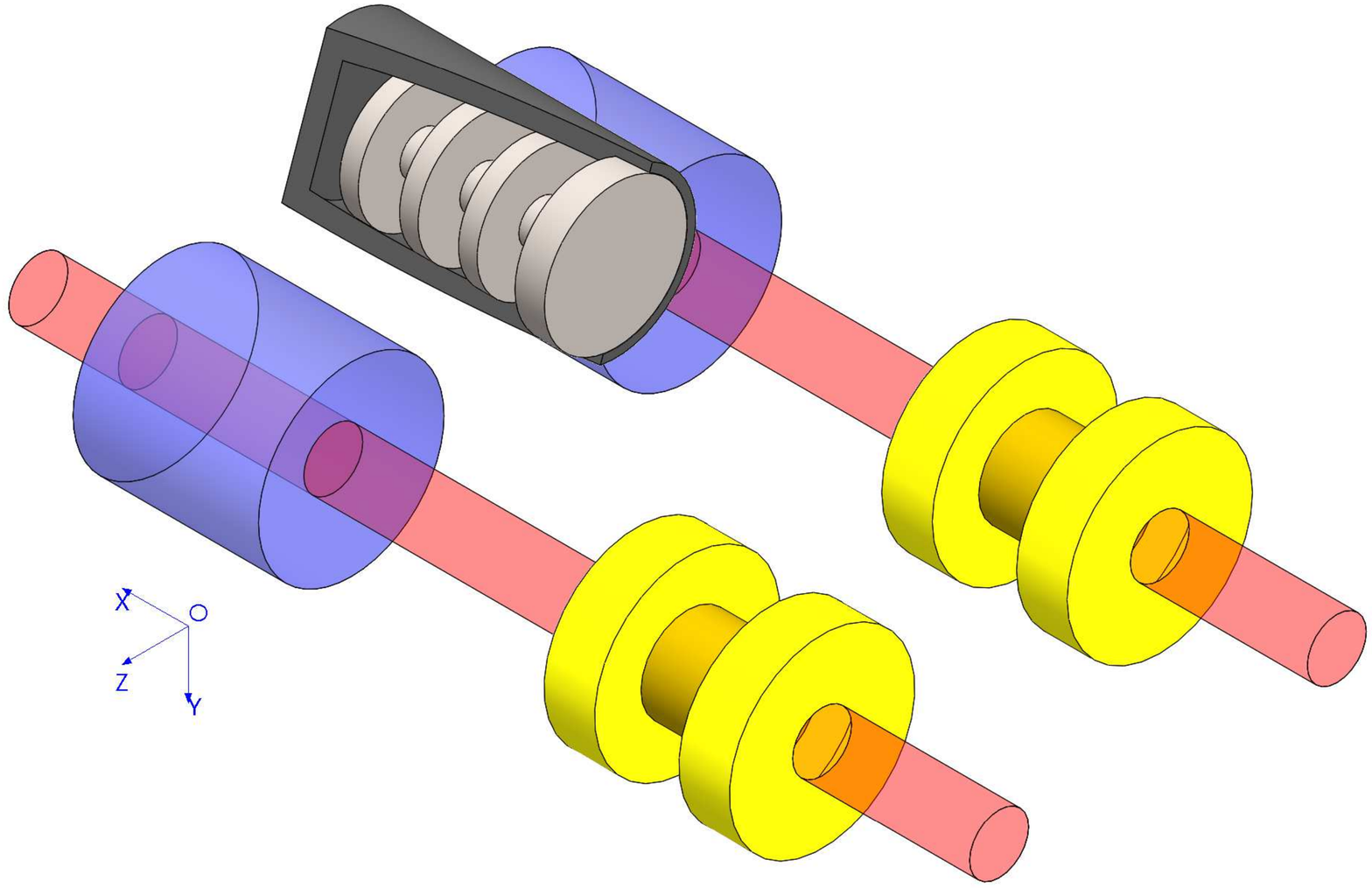}
\caption{\label{fig:setup} The laser beams (in red) propagate along the $x$ direction and cross parallelly two atomic cells (in blue). A static magnetic field is oriented along $z$ and varies with the position $x$ to the purpose of performing frequency encoding. Electromagnets (in yellow) produce a strong \textit{dressing} field oscillating along $x$ much faster than the atomic Larmor frequency. The dressing field has a position-dependent intensity, such to restore the AMR width. The NMR sample cartridge (in dark and light grey) is represented sectioned to show the internal structure.
Both the cartridge and the cells are merged in the inhomogeneous field required for the MRI frequency encoding (typical values are $B_z= 4 \mu$T and $\partial B_z/\partial x = 40$~nT/cm), and the magnetometer sensitivity is restored by the inhomogeneous dressing.
}
\end{figure}

A synthetic description of the apparatus is reported here below, referring to pertinent bibliography for its specific parts.
The core of the setup (see Fig.\ref{fig:setup}) is a dual channel OAM  working in a Bell and Bloom configuration \cite{biancalana_apb_16}.

The strength and the first-order gradient of the field in which the OAM operates are numerically controlled and optimized. Namely, eight numerically controlled current sources \cite{biancalana_rsi_17} supply the field and field-gradient coils. Automated procedures enable the nulling of the gradient terms and guarantee the field alignment along a fixed direction.

The dual sensor detector produces two magnetometric signals which contain the measurement of field variations due to both far-located and close-located sources. The first contribution is dominant and appears with the same sign on the two sensors. The second term appears with opposite signs, provided that the close-located source is opportunely displaced with respect to the sensors. 
 
The two signals are recombined to extract their common mode (CM) term and difference-mode (DM) term: far-located sources (which in our measurement constitute a disturbance) contribute to the CM only, while the MRI signal appears in the DM one.
 
Compensating the CM term has a twofold advantage: (i) the disturbances and drifts of the field which would affect the nuclear precession are removed; (ii) unavoidable imperfections of the differential system (a limited CM-rejection ratio) let the CM appear residually with the DM term, thus a preliminary reduction of the CM term improves the DM signal-to-noise ratio.
 
To this aim, while  extracting  the MRI signal from the DM term, the CM term is used  to feed a self-optimized closed loop system \cite{biancalana_FPGAstab_prappl_19} to actively compensate the external disturbances, which are strongly present due to the unshielded nature of the set-up.

As represented in Fig.\ref{fig:setup}, both the Cs atoms and the sample protons precess around a static (stabilized) field $B_z$ oriented transversely with respect to the beam propagation axis ($x$).
The field $B_z$ is made dependent on the position  $x$ to the purpose of performing MRI frequency-encoding. Its  gradient $G=\partial B_z/\partial x$ is set by permanent magnets  arranged  in a quadrupolar configuration: $B_z=B_0+Gx$, where $B_0$ is the field at the center of the cell.  
The proton and Cs Larmor frequencies set by $B_z$ are thus position dependent  along the optical axis $x$.
Typical values of G amount to tens of nT/cm, which would broaden the AMR width from few tens up to several hundred Hz, degrading and eventually destroying the OAM operativity.

Based on the IDEA method described in Ref.\cite{biancalana_IDEA_prappl_19},  an inhomogeneous magnetic dressing technique is applied to recover the OAM operativity.
The presence of a dressing field, makes the time evolution of the atomic magnetization $\vec M$ more complicated than a simple precession around the static field $B_z$. In particular, a strong $B_D$ field oscillating along $x$ at a frequency $f$ much larger than $\gamma_{\mathrm{Cs}} B_z$ (let $\gamma_{\mathrm{Cs}}=2 \pi 3.5~\mathrm{Hz/nT}$ be the gyromagnetic factor of Cs ground state
) makes the $M_y$ and $M_z$ components follow a deeply modified trajectory on the Bloch sphere. 

In contrast, the $M_x$ component --the polarimetrically measured quantity that provides the magnetometer output-- keeps oscillating harmonically \cite{  bevilacqua_pra_12}. The effect of $B_D$ on the $M_x$ evolution is just a reduction of its oscillation frequency. The reduced frequency $\nu_D$ depends on the strength of the dressing field $B_D$ and on $f$,  according to

\begin{equation}
\nu_D=J_0\left (\frac{\gamma_{\mathrm{Cs}} B_D }{2\pi f} \right ) \nu_0,
\label{eq:nuD}
\end{equation}
where $J_i$ is the $i$th order Bessel function of the first kind and $\nu_0=\gamma_{\mathrm{Cs}} B_z/2\pi$ is the precession frequency in absence of dressing field \cite{haroche_prl_70}.

An inhomogeneous dressing can compensate for the detrimental effects of the field gradient used for MRI frequency encoding.  
As described in Ref.\cite{biancalana_IDEA_prappl_19}, if the  strength of $B_D$  has an appropriate dependence on $x$, the dressing can  compensate  the position-dependent AMR frequency shift caused by the gradient of $B_z$ and the AMR width can be restored.

To this end, each  sensor is coupled to an electromagnetic $B_D$ source (a coil wound on a hollow-cylinder ferrite, with the laser beams passing across the hole). 

In a dipole 
approximation,  
at a distance $x$ from the center of the cell, the dressing field $B_D$ is
\begin{equation}
 \label{eq:BD}
 B_D(x, t) =\frac{\mu_0}{2\pi}\frac{m(t)}{(x_0+x)^3} = B_{D0}(x) \cos (2\pi f t),
\end{equation}
where $\mu_0$  is the vacuum permittivity, $m(t)=m_0  \cos (2\pi f t)$
is  the oscillating  dipole momentum, and $x_0$  is the  distance of  the dipole from the cell center.  

Taking into account the dependence on $x$ of both the static and the dressing fields,   $M_x$ oscillates harmonically at a frequency
\begin{equation}
\label{eq:nuDx}
\nu_{D}(x)=\frac{\gamma_{\mathrm{Cs}}}{2 \pi}\left( B_0 + Gx\right) J_0 \left( \frac{\gamma_{\mathrm{Cs}} B_{D0}(x)}{2\pi f }\right),
\end{equation}
in a first-order Taylor approximation, 
\begin{equation*}
  \label{eq:eq:omega:svil}
  \begin{split}
  \nu_D(x) & = \nu_D(0) + \nu_D^{\prime}(0) \, x + O(x^2)\\ 
  & \approx \frac{\gamma_{\mathrm{Cs}}}{2\pi} \left( B_0 J_0(\alpha) +  \left[ \frac{3 B_0 \alpha
  J_1(\alpha)}{x_0} + G  J_0(\alpha) \right] x \right ),\\
  \end{split}
\end{equation*}
where $\alpha = (\mu_0/4\pi^2) (\gamma_{\mathrm{Cs}} m_0)/(f x_0^3) $, so that the condition for compensating the effect of the gradient $G$ reduces to
\begin{equation}
  \label{eq:def:grad}
   -3 \frac{B_0}{x_0} \frac{\alpha J_1(\alpha)}{J_0(\alpha)} = G. 
\end{equation}

Under this condition,  the OAM performance is recovered, so to guarantee the sensitivity necessary to detect MRI signals. It is worth noting that the dressing effect is negligible for protons, because of their much smaller gyromagnetic factor: $\gamma_{\mathrm{H}} \ll \gamma_{\mathrm{Cs}}$, which makes  $J_0 (\gamma_{\mathrm{H}} B_D /2\pi f ) \approx 1$.

Differing from the case studied in Ref.\cite{biancalana_IDEA_prappl_19}, here two distinct Cs cells are used, each of them equipped with a dressing dipole: an arrangement that increases the NMR signal improving the MRI performance. 
With this feature, the present results demonstrate that a multiple-sensor arrangement can be built, with one oscillating IDEA dipole for each sensor. It is possible to maintain  cross-talking between dressed sensors at a negligible level. 

\begin{figure}[ht]
\includegraphics[angle=0, width=\columnwidth]{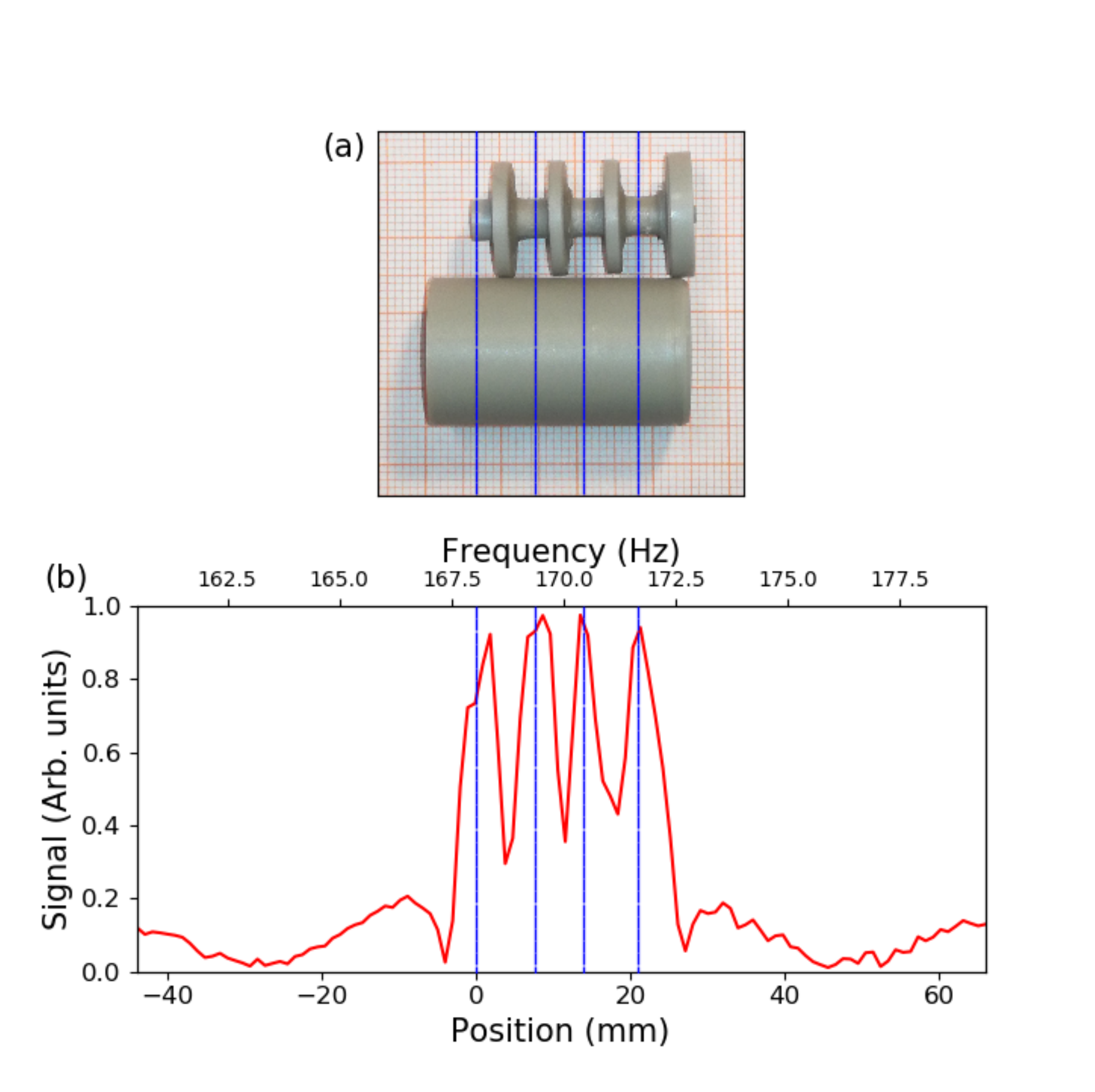}
\caption{\label{fig:mriefoto} The panel (a) shows a photograph of the open cartridge with its internal structure. In the panel (b) an uni-dimensional image is shown as reconstructed from an average trace corresponding to tight (0.08~mm interval) positioning of the sample. Both (a) and (b) are in the same scale, blue lines associate MRI peaks to their origin in cartridge volume.
}
\end{figure}

The NMR sample is made of water protons contained in a polymeric cartridge having the structured shape shown in Figs.\ref{fig:setup} and \ref{fig:mriefoto}(a): it
is a cylinder --19~mm  in diameter, 32~mm  in length-- that
contains three disks  --2~mm in thickness-- separated by 5~mm from each other, the water (in hydrogel) is confined in the four complementary disks. 
Care is taken to avoid ferromagnetic contamination of the container \cite{biancalana_hybrid_rsi_19}.

The setup includes an Halbach permanent-magnet assembly to premagnetize the sample at 1~T and a pneumatic shuttle system \cite{biancalana_rsi_14} to move it cyclically to the measurement region 
(see Refs.\cite{ biancalana_zulfJcoupling_jmr_16, biancalana_DH_jpcl_17} for additional details).

\begin{figure*}[ht]
\includegraphics[angle=0, width= 2  \columnwidth]{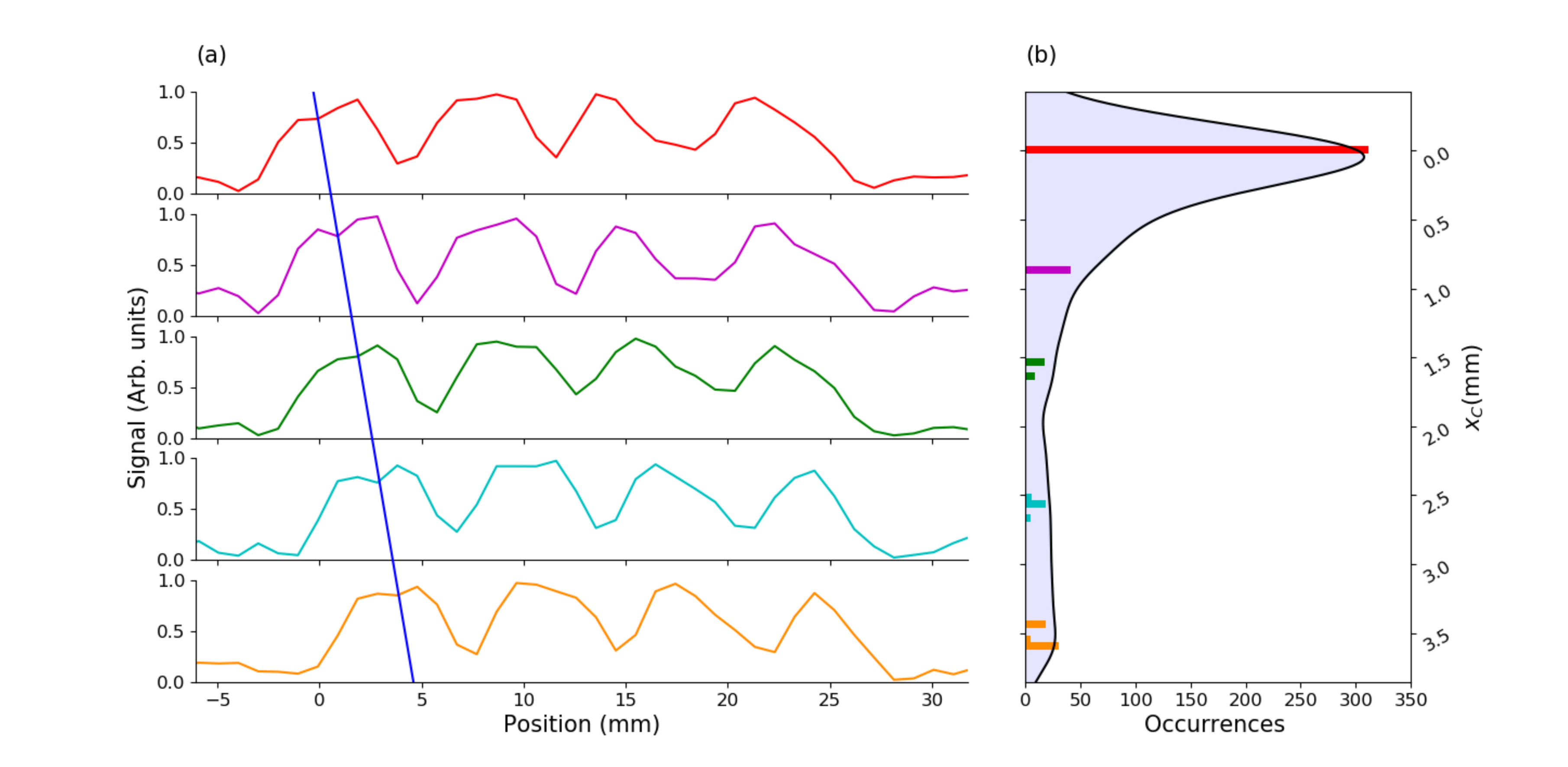}
\caption{\label{fig:bella} The histogram (b) shows the distribution of 
$x_C$
over a large set of measurements. Five subsets of measurements 
corresponding to narrow intervals of $x_C$
have been selected to reconstruct the MRI traces shown in the panel (a). To this end, the selected time traces are averaged and analysed by FFT: the real part of the Fourier transform is shown. The inclined blue line guides the eyes to localize the leftmost peak position of each image as it varies along the histogram (3.5~mm from the upper to the lower plot). The maxima displacements match the corresponding camera estimations $x_C$ with submillimetric precision.}
\end{figure*}

The cartridge position along $x$ slightly changes shot-by-shot. At each measurement, a camera monitors such sample positioning  with respect to the sensors. An automated image analysis provides a localization  $x_C$ (with respect to a fixed origin of the camera abscissa) with an uncertainty of 0.08~mm. The  $x_C$ data are registered together with the corresponding NMR traces to be used in post processing.

Correspondingly to the cartridge internal structure shown in Fig.\ref{fig:mriefoto}(a), a 1-D MRI profile is reported in Fig.\ref{fig:mriefoto}(b). That profile is obtained by averaging over 300 traces.
The latter are selected on the basis of the measured  $x_C$ values, which in this case fall within a 0.08~mm interval. The 1-D image 
clearly shows the four peaks corresponding to the hydrogel disks. Such plot is directly obtained as real part of the average-trace Fourier transform: just minor 
additional data manipulation is performed.
The spatial resolution is set by the NMR intrinsic and instrumental linewidth, which --in absence of field gradient-- amounts to 1.5~Hz: its effect is only partially compensated in the spectral analyses used in this paper.

For an alternative  evaluation of  the  MRI spatial resolution, we use a set of traces collected in more than 2000 shots.  
Subsets of traces corresponding to narrow $x_C$ intervals in the statistic distribution of $x_C$ are averaged to produce MRI profiles. The profile displacements are then compared with the $x_C$ variations.

In Fig.\ref{fig:bella}, the  panel (b) shows the histogram of $x_C$ positions and five subsets extracted to produce the average time-traces that finally result in the MRI profiles shown in (a). The four-peaks profile shifts progressively in accordance with the $x_C$, with submillimetric accuracy. 

In conclusion, this work demonstrates that the IDEA method enables the construction of submillimetric ULF-MRI setups with a dual-sensor NMR detection, based on Bell and Bloom OAMs, even operating in an unshielded environment. 
Despite their lower SNR compared to SQUIDs, the OAMs bring the advantage of an intrinsic robustness with respect to the temporary application of strong magnetic fields, such as those necessary for spin premagnetization and manipulation. 
The described setup does not include ferromagnetic shielding, which would facilitate a rearrangement devoted to apply a strong premagnetization field in the proximity of the sensors, so to build a fully static, \textit{in-situ} experiment. The achieved sensitivity and the residual magnetic noise make our sensors about one order of magnitude less performing compared to typical SQUID and SERF-OAM devices: an aspect which would require progresses in view of practical applications. Further steps of our research foresee further improvements of the active field compensation system and the implementation of a in-loco premagnetization coil.
In the perspective of real-world applications, it is worth mentioning that our two-sensor arrangement demonstrates the possibility of extending the method to multi-sensor setups: more complex sensor arrays can exploit the  IDEA concept, and have the potential of improving the sensitivity, accelerating the measurement and enabling ULF-MRI of larger-size samples.

\section*{Supplementary Material}
See supplementary for experimental details and for a concise description
of the various elements of the setup and their principle of operation

\bibliography{biblio_mm}



\vfill
\newpage
\section{supplemental material}

\subsection{Setup overview}
The setup is made up of two independent parts, an MRI subsystem and a magnetometric sensor recording MRI signal.
The sensor is a Bell and Bloom \cite{bellandbloom_prl_61}  OAM, which operates in an unshielded environment \cite{biancalana_apb_16}, and has a dual channel head allowing for differential measurements.

Due to the presence of static bias field (along $z$) the sensors, being scalar in nature, have first-order response only to $B_z$ variations: 
\begin{equation}
    \delta B= | \vec B_0 - \delta \vec B| \approx \frac{\vec B_0 \cdot \delta \vec B}{ B_0} = \delta B_{\parallel}.
        \label{eq:deltaB}
\end{equation}

The common mode variations of this field component is actively stabilized.

Additional coils, belonging to MRI subsystem, permit the application of resonant or non resonant (non-adiabatically varying) field pulses, to manipulate (tipping) the nuclear spins \cite{biancalana_zulfJcoupling_jmr_16, biancalana_DH_jpcl_17}. The magnetometric operation is destroyed by these pulses, and recovers completely in about 100~ms.

In the present implementation, a remote magnetization approach is applied. To perform MRI, a pneumatic system moves cyclically the sample from a 1~T premagnetizing assembly, configured as an Halbach  magnet array to the measurement region and backwards \cite{biancalana_rsi_14}. The sample transfer, tipping pulse, and recovery time amount globally to about 250ms: this interval constitutes the dead-time between the premagnetization and the measurement.

A constant field gradient is applied to the measurement region with the aim of performing frequency encoding necessary to the MRI. A good spatial resolution requires a rather strong gradient (30-100 nT/cm). The AMR resonance would be completely destroyed by this gradient, with a width increase from about 25~Hz up  to 200-700~Hz. A technique based on inhomogeneous magnetic dressing (IDEA method \cite{biancalana_IDEA_prappl_19})  permits to maintain the AOM operability despite the gradient, so to register successfully MRI profiles.

\subsection{Magnetometer}
The OAM uses two sealed cells (cylindrical shape, 2 cm in length, 2.2 cm in diameter) with Cs vapour and buffer gas, and two fiber-coupled laser sources whose radiation is mixed, split and delivered to both the cells (see Fig.1). The first laser radiation is at milli-Watt level and optically pumps the Cs vapor  into a stretched  state, which then evolves due to the presence of transverse magnetic field. This pump radiation is tuned to the proximity the Cs $D_1$ line (894 nm), and its optical frequency is periodically made  resonant with the $F_g=3 \rightarrow F_e=4$  transition of the $D_1$ line set. This condition  produces both light-narrowing \cite{sch_pra_11, biancalana_pra_16} --due to strong hyperfine pumping to the $F_g=4$ ground state-- and  Zeeman pumping, due to the weak excitation of far detuned resonances starting from the $F_g=4$ level, and to the light polarization, which is made circular. 

The second laser radiation is attenuated down to microwatt level, and serves as a probe. It co-propagates with the pump, and is linearly polarized. The rotation of the probe polarization plane is a measure of the atomic vapour magnetization along the propagation direction (let it be $x$). The probe beam is tuned to the proximity
of the $F_g=4$ manyfold in the $D_2$ line, at 852 nm. 

The wavelength difference of the two radiations facilitates two tasks: it is possible to use a multiorder waveplate which acts as a $\lambda/4$  plate on the probe and as a full-$\lambda$ on the pump radiation, to obtain the aforementioned polarizations; it is possible to filter out the pump radiation after the interaction with Cs, to perform polarimetry of the probe radiation only.

The pump radiation is made resonant periodically, and sinchronously with the atomic precession   around a transverse ($z$ oriented) static magnetic field. 

Let $\omega_M$ and  $\omega_L$ be the angular frequency of the laser modulation signal and of the atomic precession, respectively.
Scanning of $\omega_M$ around $\omega_L$ enables the registration of the AMR profile. Linewidths as narrow as 25~Hz are recorded in operative conditions. 
This width is mainly due to spin-exchange relaxation and to probe perturbation: a resonance narrowing down to 5~Hz is observed when operating at lower cell temperature and weaker (or more detuned) probe radiation. 

To the end of maintaining a high sensitivity level, it is important to preserve a narrow resonance width. The presence of buffer gas (3 kPa N$_2$) prevents time-of-flight line broadening and radiation trapping phenomena. Spurious field gradients are accurately compensated, and the IDEA method is used to counteract the effect of the strong field gradient deliberately applied in MRI to perform the frequency encoding .

During the measurements, $\omega_M$ is made resonant ($\omega_M=\omega_L$) and kept constant.
In addition, the modulation signal at $\omega_M$ is superimposed to an adjustable offset, by which the optical frequency of the pump laser is kept at its optimal value \cite{biancalana_apb_16}.

\subsection{IDEA (principle of operation)}

In the proximity of each sensor, displaced about 10 cm along the $x$ direction, a hollow ferrite nucleus (with the laser beam passing through the hole) wound by a solenoidal coil  produces a dipolar field used to dress magnetically the Cs atoms.

Such dipolar  oscillating  field $B_D$ is oriented along $x$ --thus transversely to the static field $B_z$-- and has an important inhomogeneity along the $x$ direction.

The static field $B_z$ is made not-homogeneous along $x$ for the MRI-frequency encoding purpose. Its  gradient $G=\partial B_z/\partial x$ is set by permanent magnets (small -- 4 cm$^3$ -- Nd magnets, well displaced --about 1 m-- from the measurement region) and arranged  in a quadrupolar configuration: both the sample nuclei and the sensor atoms precess in a field whose static $z$ component is well described by $B_z=B_0+Gx$, where $x$ is the distance from the sensor center, and $B_0$ is the field at the center of the cell.  
Thus the Cs and proton Larmor frequencies set by $B_z$ are position dependent along the optical axis $x$.

The dressing  field $B_D$ (Eq.2) is produced by an alternating current supplying
a coil (100 turns wound on a hollow cylinder ferrite, Richco RRH-285-138-286: 28.5 mm external diameter, 13.77 internal diameter, 28.57 mm length). This device produces a dipolar field with adjustable intensity (several $\mu$T at 40 kHz). The inductance is measured (1.35 mH) and a resonant LC circuit is built to enhance $B_D$.
 
The presence of the dressing field, makes the time evolution of the atomic magnetization more complicated than a simple precession around the static field $B_z$. In particular, the $y$ and $z$ components of the atomic magnetization follow a deeply modified trajectory on the Bloch sphere. 
On the contrary, its $x$ component keeps oscillating harmonically. The only effect of $B_D$ on the evolution of the $x$ component is slowing down of its oscillation frequency.
 
In conclusion, the  $x$ component of the atomic magnetization --the observed quantity revealed by the Faraday rotation of the probe beam polarization-- oscillates
harmonically at a dressed  (reduced) angular frequency with respect to
its    unperturbed     precession    around    the     static    field
\cite{bevilacqua_pra_12} (see Eq.1).

\subsection{Magnetic Field sources}
To adjust the static field and its inhomogeneties, the system includes three large size (1.8 m side) square Helmholtz coils to compensate the environmental field, two anti-Helmholtz coils (same size) to compensate the diagonal elements of the field Jacobian, and three pairs of dipoles arranged to produce quadrupolar fields in the measurement region, which compensate the off-diagonal elements. The latter are made of permanent magnets for coarse compensation and numerically controlled electromagnets for the fine adjustments.
Eight numerically controlled current sources designed with a hybrid (switching plus linear architecture) \cite{biancalana_rsi_17} supply the field and field-gradient coils. Automated procedures are applied to null the gradient terms and the transverse field, as well as to set the amplitude of the residual field along the $z$ direction. The latter is furtherly stabilized as discussed in section \ref{subsec:ML}.

For nuclear spin manipulation, a smaller (50 cm side) Helmholtz pair is used to apply appropriate pulses ($\pi/2$ tipping pulses).
The permanent magnets necessary to generate the quadrupole field with the frequency-encoding gradient G are inserted after the completion of the above described field optimization procedure.

\subsection{Magnetic Field stabilization}
\label{subsec:ML}

As said in the main text, the dual sensor detector produces two magnetometric signals which contain the measurement of field variations due to both far-located and close-located sources. The first contribution is dominant and appears with the same sign on the two sensors. The second term appears with opposite signs, provided that the close-located source (that is the sample) is opportunely located with respect to the sensors. 

It is useful to recombine the two signals to extract their common mode (CM) term and difference-mode (DM) term: far-located sources (which in our measurement constitute a disturbance) contribute to the CM only, while the MRI signal appears in the DM one.

The polarimetric signal of one channel is modeled as
\begin{equation}
s(t)=A e^{i[\omega_M t + \varphi(t)]} .
\label{eq:segnale}   
\end{equation}

The measured polarimetric signal is acquired, real-time numerically demodulated and elaborated by a field-programmable-gate-array (FPGA). 
After the demodulation, $\varphi(t)$ is converted into an error signal by means of an IIR filter characterized by a set of forward and backwards coefficients. Afterwards, the error signal drives the current generator supplying compensation coils.

As discussed in the appendix of Ref.\cite{biancalana_apb_16}, a small variation $\delta B(t)$ of the field component along the static field $B_z$ acts on  $\varphi (t)$ according to the equation 
\begin{equation}
  \label{eq:fin:phi:lin}
  \dot{\varphi} =   - \gamma_{\mathrm{Cs}}\delta B(t) - \Gamma \varphi, 
\end{equation}
where $\Gamma$ is the AMR linewidth.

IIR forward and backwards coefficients are initially chosen in order to combine the phase with its derivative, calculated with a finite-difference extimation.
Thus, filtering the phase $\varphi(t)$ by means of IIR filter, ensures that the resulting error-signal is proportional to magnetic field variation as in Eq.\ref{eq:fin:phi:lin}.
Thereafter IIR coefficients are automatically optimized using the Nelder-Mead algorithm; in such way the evaluation of error signal considers firstly the ideal magnetometer response, but is secondly adapted to compensate delays occurring during elaboration process and specific behaviour of the chain of elements that constitute the loop. 
The optimization process is based on the analysis of the polarimetric signal power spectrum. More precisely, the IIR optimization is targeted to maximize the energy associated to the carrier peak at $\omega_M$ with respect to other spectral components due to disturbances.

The error signal at the output of the IIR filter feeds a voltage-to-current converter, which  supplies a pair of compensation Helmholtz coils. The magnetic field generated by means of the coils is nominally oppostite to the CM of the magnetic disturbances detected by the sensors, in such way those disturbances can be successfully reduced to a large extent \cite{biancalana_FPGAstab_prappl_19}.

\subsection{Polarimetric signal elaboration}
\label{subsec:signelab}
After the interaction with each atomic sample, the pump radiation is stopped by a narrowband interference filter and the probe beam polarization is analyzed by a balanced polarimeter (a 45 degree Wollaston prism followed by a couple of Si photodetectors, whose photocurrent unbalance is preamplified by a ultra-low noise transimpendance amplifier).

The two polarimetric signals (one for each cell) are nearly harmonic (Eq.\ref{eq:segnale}), at the oscillation frequency of the $x$ component of the atomic magnetization.

These signals are digitized by a 1MS/s 16 bit card  for the duration of the measurement: 6 sec after the application of the tipping pulse, with a start delay of 150 msec, necessary to recover the magnetometer's steady state. This signal digitization is performed parallelly and independently from the FPGA real-time acquisition discussed in Sec.\ref{subsec:ML}.

Provided that the pump modulation is performed at a fixed frequency matching the AMR center, the measured signal oscillates at such forcing frequency, while the field information is contained in its phase $\varphi$, as mentioned above (Eq.\ref{eq:fin:phi:lin}).

A discrete Hilbert transform is used to infer the imaginary part of the digitized polarimetric signal. The phase $\varphi$ is then extracted from the complex signal after having demodulated the frequency of the forcing term, its time derivative is finally evaluated so to estimate the field variation at the time  as $\delta B (t)= (\Gamma \varphi - \dot \varphi)/ \gamma_{\mathrm{Cs}}$. 

The mean value of the two $\delta B$'s measured by the two sensors constitutes the CM term (driven by disturbances from far locates field sources), while their difference (DM term) contains information about close located  sources and is further elaborated to extract the MRI traces.

\subsection{MRI detection and signal elaboration}
As shown in Fig.1 the MRI sample is located at the same $x$, intermediate $z$ and different $y$ with respect to the sensors. Consequently, when the nuclear magnetization precesses around the static field $B_z$, it produces periodic variations of the measured field component (the measured $\delta B$ is the variation of the $z$ component, Eq.\ref{eq:deltaB}). These variations occur with opposite signs in the two sensors, so to appear in the DM term.

The reconstructed images reported in this work are obtained from the real part of the Fourier transform of the DM term $S(t)$: $\hat S(\Omega)= \mathcal{F}(S(t))$, then scaling from $\Omega$ to $x$ via the gradient $\gamma_H G$.

Only minor additional signal manipulations are applied. The first one is a time-domain windowing having the dual scope of apodizing the signal, reducing the noise contribution from the final (signal-less) part of the time trace, and of partially compensating the exponential decay of the NMR signal. 

To this aim we apply a window in the form
\begin{equation}
W(t)=\exp(t/T) \left[1- \tanh \left (\beta (t-t_{\mathrm{cut}}) \right ) \right]^2,
\end{equation}
where $T$ is kept lower than the  measured proton decay time, and $t_{\mathrm{cut}}$ and $\beta$ are parameters adjusting the position and the slope of the transition to zero of the term in square brackets.

The second operation is to take into account the delay $t_0$ between the application of the tipping pulse and the measurement: the transformed signal $\hat S(\Omega)$ is scaled by $\exp(i \Omega t_0)$ before extracting the real part.

The latter operation (that is a translation in the time domain) comes with some artefacts, making a third manipulation opportune. Due to spurious deterministic disturbances at 50~Hz and harmonics as well as at $\omega_M$, additional terms appear superimposed to the MRI profile. These terms have a periodicity $t_0$ in the frequency domain, which facilitates their identification and subtraction.

\subsection{MRI Sample}
The sample container has the external shape of a cylinder (19~mm in diameter diam, 32~mm in length). It is made of polyether ether ketone (PEEK, Victrex), a non-conductive material selected for its excellent mechanical properties in matter of machinability and robustness (the cartridge must withstand thousands of shuttling stresses).

For MRI test purposes, the cap of the cartridge includes an extension (see Figs.1 and 2), which makes the measured material (water in hydrogels, to prevent diffusion processes) occupy four possible regions having the shape of hollow disks (5mm thick, 15mm external diameter, 5mm internal diameter). 

As previously pointed out \cite{biancalana_hybrid_rsi_19}, despite the negligible intrinsic magnetic remanence of the PEEK, the machining process may introduce detectable surface contamination, making the container spuriously and weakly ferromagnetic. 
It is known that a paramagnetic response  produces MRI artefacts that are important at high field, but negligible in ULF regime. On the other hand, the ferromagnetic response produce effects that persist in ULF, up to  become a dominant artefact source.  Lathing the cartridge with non magnetic tools (glass and Ti blades in our case), turned out to be necessary to prevent this problem.


\end{document}